\documentclass[preprint,superscriptaddress]{revtex4-1} 
\usepackage{float} 
\usepackage{amsmath} 
\usepackage{amssymb} 
\usepackage{longtable} 
\usepackage{graphicx} 
\usepackage{hyperref} 
\usepackage{dcolumn} 
\usepackage{mhchem} 
\usepackage{epstopdf} 
\usepackage{cleveref} 
\begin{document} 

\title{Assessment of the LFAs-PBE exchange-correlation potential for high-order harmonic generation of aligned $\text{H}_{2}^{+}$ molecules} 

\author{Hsiao-Ling Sun} 
\affiliation{Department of Physics, National Taiwan University, Taipei 10617, Taiwan} 
\affiliation{Physics Division, National Center for Theoretical Sciences (North), National Taiwan University, Taipei 10617, Taiwan} 

\author{Wei-Tao Peng} 
\affiliation{Department of Physics, National Taiwan University, Taipei 10617, Taiwan} 

\author{Jeng-Da Chai} 
\email[Author to whom correspondence should be addressed. Electronic mail: ]{jdchai@phys.ntu.edu.tw} 
\affiliation{Department of Physics, National Taiwan University, Taipei 10617, Taiwan} 
\affiliation{Physics Division, National Center for Theoretical Sciences (North), National Taiwan University, Taipei 10617, Taiwan} 
\affiliation{Center for Theoretical Sciences and Center for Quantum Science and Engineering, National Taiwan University, Taipei 10617, Taiwan} 

\date{\today} 

\begin{abstract} 

We examine the performance of our recently developed LFAs-PBE exchange-correlation (XC) potential [C.-R. Pan, P.-T. Fang, and J.-D. Chai, {\it Phys. Rev. A}, 2013, {\bf 87}, 052510] for the 
high-order harmonic generation (HHG) spectra and related properties of $\text{H}_{2}^{+}$ molecules aligned parallel and perpendicular to the polarization of an intense linearly polarized laser 
pulse, employing the real-time formulation of time-dependent density functional theory (RT-TDDFT). The results are compared with the exact solutions of the time-dependent Schr\"odinger 
equation as well as those obtained with other XC potentials in RT-TDDFT. Owing to its correct $(-1/r)$ asymptote, the LFAs-PBE potential significantly outperforms conventional XC potentials 
for the HHG spectra and the properties that are sensitive to the XC potential asymptote. Accordingly, the LFAs-PBE potential, which has a computational cost similar to that of the popular 
Perdew-Burke-Ernzerhof (PBE) potential, can be very promising for the study of the ground-state, excited-state, and time-dependent properties of large electronic systems, extending the 
applicability of density functional methods for a diverse range of applications. 

\end{abstract} 

\maketitle

\section{Introduction} 

Due to its reasonable accuracy and computational efficiency, time-dependent density functional theory (TDDFT) \cite{RG} has been one of the most popular methods for studying the excited-state 
and time-dependent properties of large electronic systems \cite{Gross,TDDFT,TDDFT2}. Nevertheless, as the exact time-dependent exchange-correlation (XC) potential $v_{xc}({\bf r},t)$ in TDDFT 
has not been known, an approximate treatment for $v_{xc}({\bf r},t)$ is necessary for practical applications. 

For a system under the influence of a slowly varying external potential, the best known approximation for $v_{xc}({\bf r},t)$ is the adiabatic approximation: 
\begin{equation} 
v_{xc}({\bf r},t) \approx \frac{\delta E_{xc}[\rho]}{\delta \rho({\bf r})}\bigg|_{\rho({\bf r})=\rho({\bf r},t)}, 
\label{eq:vxc} 
\end{equation} 
where $v_{xc}({\bf r},t)$ is approximately given by the static XC potential (i.e., the functional derivative of the XC energy functional $E_{xc}[\rho]$) evaluated at the instantaneous density 
$\rho({\bf r},t)$. Although the entire history of the density (i.e., memory effects) is ignored in the adiabatic approximation, the accuracy of the adiabatic approximation can be surprisingly high in 
many situations (even if the system is not in this slowly varying regime). However, since the exact $E_{xc}[\rho]$, the essential ingredient of both Kohn-Sham density functional theory 
(KS-DFT) \cite{HK,KS} (for ground-state properties) and adiabatic TDDFT (for excited-state and time-dependent properties), remains unknown, there has been considerable effort invested in 
developing accurate density functional approximations (DFAs) for $E_{xc}[\rho]$ to improve the accuracy of both KS-DFT and adiabatic TDDFT for a wide range of 
applications \cite{Gross,TDDFT,TDDFT2,DFTreview,DFTreview2,DFTreview3}. 

Conventional density functionals (i.e., semilocal density functionals) are based on the local density approximation (LDA) \cite{LDAX,LDAC}, generalized gradient approximations (GGAs), and 
meta-GGAs (MGGAs) \cite{ladder}. They are computationally efficient for large systems, and reasonably accurate for properties governed by short-range XC effects, such as low-lying valence 
excitation energies. However, they can predict qualitatively incorrect results in situations where an accurate description of nonlocal XC effects is 
critical \cite{Gross,TDDFT,TDDFT2,DFTreview,DFTreview2,DFTreview3}. For example, the LDA and most GGA XC potentials do not exhibit the correct asymptotic behavior in the asymptotic region 
($r \rightarrow \infty$) of a molecular system, yielding erroneous results for the highest occupied molecular orbital (HOMO) energies, high-lying Rydberg excitation energies \cite{R9,R10,R11,R12}, 
charge-transfer (CT) excitation energies \cite{R12,R15,R16,R17,R18,R19,Peng}, and excitation energies in completely symmetrical systems where no net CT occurs \cite{R21}. 

To properly describe properties that are sensitive to the asymptote of the XC potential, two different density functional methods with correct asymptotic behavior have been actively developed over 
the past two decades. One is the long-range corrected (LC) hybrid scheme \cite{R23,R24,BNL,wB97X,wB97X-D,wM05-D,LC-D3,op,wB97X-2,LCHirao,LC-wPBE,terf_JAP,LC-DFT,LCgauHirao}, 
and the other is the asymptotically corrected (AC) model potential scheme \cite{LB94,LBa,AA,AC1,Tozer,LFA,AK13,AC_Trickey,AC_Truhlar}, both of which have recently become very popular. In 
the LC hybrid scheme, as 100\% Hartree-Fock (HF) exchange is adopted for long-range interelectron interaction, an AC XC potential can be naturally generated by the optimized effective potential 
(OEP) method \cite{OEP1,OEP2,OEP3,DFTreview}. However, for most practical calculations, the generalized Kohn-Sham (GKS) method, which uses orbital-specific XC potentials, has been 
frequently adopted in the LC hybrid scheme to avoid the computational complexity involved in solving the OEP equation, as the electron density, total energy, and HOMO energy obtained with the 
GKS method are very close to those obtained with the OEP method \cite{OEP3,DFTreview}. Recently, we have examined the performance of various XC energy functionals for a diverse range of 
applications \cite{LCAC,EB}. In particular, we have shown that LC hybrid functionals can be reliably accurate for several types of excitation energies, involving valence, Rydberg, and CT excitation 
energies, using the frequency-domain formulation of linear-response TDDFT (LR-TDDFT) \cite{R4,LR-TDDFT}. Nevertheless, due to the inclusion of long-range HF exchange, the LC hybrid scheme 
can be computationally expensive for large systems. 

By contrast, in the AC model potential scheme, an AC XC potential is modeled explicitly with the electron density, retaining a computational cost comparable to that of the efficient semilocal density 
functional methods. In the asymptotic region of a molecule where the electron density decays exponentially, the popular LB94 \cite{LB94} and LB$\alpha$ \cite{LBa} potentials exhibit the correct 
$(-1/r)$ decay by construction. However, as most AC model potentials, including the LB94 and LB$\alpha$ potentials, are {\it not} functional derivatives \cite{Staroverov,Staroverov2b}, the associated 
XC energies and kernels (i.e., the second functional derivatives of the XC functionals) are not properly defined. Due to the lack of an analytical expression for $E_{xc}[\rho]$, the XC energy 
associated with an AC model potential is usually evaluated by the popular Levy-Perdew virial relation \cite{virial}, which has, however, been shown to exhibit severe errors in the calculated 
ground-state energies and related properties \cite{LCAC,LFA}. In addition, due to the lack of a self-consistent adiabatic XC kernel, an adiabatic LDA or GGA XC kernel has been constantly employed 
for the LR-TDDFT calculations using the AC model potential scheme. Such combined approaches have been shown to be accurate for both valence and Rydberg excitations, but inaccurate for CT 
excitations \cite{LCAC,R20,R22new,new1,LFA}, due to the lack of a space- and frequency-dependent discontinuity in the adiabatic LDA or GGA kernel adopted in LR-TDDFT \cite{R22}. Besides, 
due to the lack of the step and peak structure in most adiabatic AC model potentials (e.g., LB94) \cite{R38}, we have recently shown that adiabatic AC model potentials can fail to describe CT-like 
excitations \cite{Peng}, even in the real-time formulation of TDDFT (RT-TDDFT) \cite{TDDFT2}, where the knowledge of the XC kernel is not needed. 

Very recently, we have developed the localized Fermi-Amaldi (LFA) scheme \cite{LFA}, wherein an exchange energy functional whose functional derivative has the correct $(-1/r)$ asymptote can be 
directly added to any semilocal functional. When the Perdew-Burke-Ernzerhof (PBE) functional \cite{PBE} (a very popular GGA) is adopted as the parent functional in the LFA scheme, the resulting 
LFA-PBE functional, whose functional derivative (i.e., the LFA-PBE XC potential) has the correct $(-1/r)$ asymptote. Among existing AC model potentials, the LFA-PBE XC potential is one of the very 
few XC potentials that are functional derivatives \cite{LFA,AK13,AC_Trickey,AC_Truhlar}. 

In addition, without loss of much accuracy, the LFAs-PBE XC potential (an approximation to the LFA-PBE XC potential) \cite{LFA} has been developed for the efficient treatment of large systems. 
For a molecular system, the LFAs-PBE XC potential can be expressed as 
\begin{equation} 
v^{\text{LFAs-PBE}}_{xc}(\textbf{r}) = v^{\text{PBE}}_{xc}(\textbf{r}) + v^{\text{LFAs}}_{x}(\textbf{r}). 
\label{eq:LFAs-PBE} 
\end{equation} 
Here $v^{\text{PBE}}_{xc}(\textbf{r})$ is the PBE XC potential, and $v^{\text{LFAs}}_{x}(\textbf{r})$ is the LFAs exchange potential: 
\begin{equation} 
v^{\text{LFAs}}_{x}(\textbf{r}) = -\sum_{A} w_{A}(\textbf{r}) \frac{\text{erf}(\omega \left|{\bf r}-{\bf R}_{A}\right|)}{\left|{\bf r}-{\bf R}_{A}\right|}, 
\label{eq:LFAs} 
\end{equation} 
where the range-separation parameter $\omega = 0.15$ bohr$^{-1}$ was determined by fitting the minus HOMO energies of 18 atoms and 113 molecules in the IP131 database to the corresponding 
experimental ionization potentials \cite{wM05-D}. In Eq.\ (\ref{eq:LFAs}), the sum is over all the atoms in the molecular system, ${\bf R}_{A}$ is the position of the atom $A$, 
and $w_{A}(\textbf{r})$ is the weight function associated with the atom $A$: 
\begin{equation} 
w_{A}(\textbf{r}) = \frac{\rho^{0}_{A}(\textbf{r})}{\sum_{B} \rho^{0}_{B}(\textbf{r})}, 
\label{eq:wA} 
\end{equation} 
ranging between 0 and 1. Here $\rho^{0}_{A}(\textbf{r})$ is the spherically averaged electron density computed for the isolated atom $A$. 
Note that even with the aforementioned approximation, the LFAs-PBE XC potential remains a functional derivative (as discussed in Ref.\ \cite{LFA}). 
Due to the sum rule of $\sum_{A} w_{A}(\textbf{r}) = 1$, the asymptote of $v^{\text{LFAs}}_{x}(\textbf{r})$ is shown to be correct: 
\begin{equation} 
\lim_{r \rightarrow \infty} v^{\text{LFAs}}_{x}(\textbf{r}) = -\sum_{A} w_{A}(\textbf{r}) \frac{1}{\left|{\bf r}\right|} = -\frac{1}{r}. 
\label{eq:LFAs_AC} 
\end{equation} 
As $v^{\text{PBE}}_{xc}(\textbf{r})$ decays much faster than $(-1/r)$ in the asymptotic region, $v^{\text{LFAs-PBE}}_{xc}(\textbf{r})$ (see Eq.\ (\ref{eq:LFAs-PBE})) retains the correct $(-1/r)$ asymptote. 

LFAs-PBE has been shown to yield accurate vertical ionization potentials and Rydberg excitation energies for a wide range of atoms and molecules, while performing similarly to their parent PBE 
semilocal functional for various properties that are insensitive to the XC potential asymptote. Relative to the existing AC model potentials that are not functional derivatives, LFAs-PBE is significantly 
superior in performance for ground-state energies and related properties \cite{LFA}. 

As LFAs-PBE is computationally efficient (e.g., similar to PBE), it may be interesting and perhaps important to understand the applicability and limitations of LFAs-PBE for a diverse range of 
applications. In this work, we examine the performance of LFAs-PBE for various time-dependent properties using adiabatic RT-TDDFT. The rest of this paper is organized as follows. In Section II, we 
describe our test sets and computational details. The time-dependent properties calculated using LFAs-PBE are compared with those obtained with other methods in Section III. Our conclusions are 
presented in Section IV.

\section{Test Sets and Computational Details} 

High-order harmonic generation (HHG) from atoms and molecules is a nonlinear optical process driven by intense laser fields 
\cite{atto,HHG,Litvinyuk,Le,Lewenstein,3steps,Chu,Chu2,Cui,Heslar,Wasserman,HHG_NP,Fowe,Fowe2,Baker,Baker2,Chu3,Chu4,Chu5,Chu6a,Chu6,Chu7,Chu8,Chu9,XChu,XChu2,HHG_N,HHGa,HHGb}. 
Recently, HHG has attracted considerable attention, as it can serve as a probe for the structure and dynamics of atoms and molecules and chemical reactions on a femtosecond time scale. Besides, 
HHG may also be adopted for producing attosecond pulse trains and individual attosecond pulses \cite{HHGa,HHGb}. In the semiclassical three-step model of HHG \cite{3steps,Lewenstein}, first, 
an electron from an atom or molecule is ionized by a strong laser field; secondly, the electron released by tunnel ionization is accelerated by the oscillating electric field of the laser pulse; thirdly, 
the electron returns and recombines with the parent ion, emitting a high-energy photon. 

In RT-TDDFT, HHG spectra can be obtained by explicitly propagating the time-dependent Kohn-Sham (TDKS) equations. As the asymptote of the XC potential should be important to the ionization, 
acceleration, and recombination steps in the HHG process, the LFAs-PBE potential is expected to provide accurate results for HHG spectra and related properties. Here we examine the performance 
of various adiabatic XC potentials in RT-TDDFT for the HHG spectra and related properties of the one-electron $\text{H}_{2}^{+}$ system, as the exact results can be easily obtained with the 
time-dependent Schr\"odinger equation (TDSE) for direct comparison. 

\begin{figure} 
\includegraphics[width=1.0\textwidth]{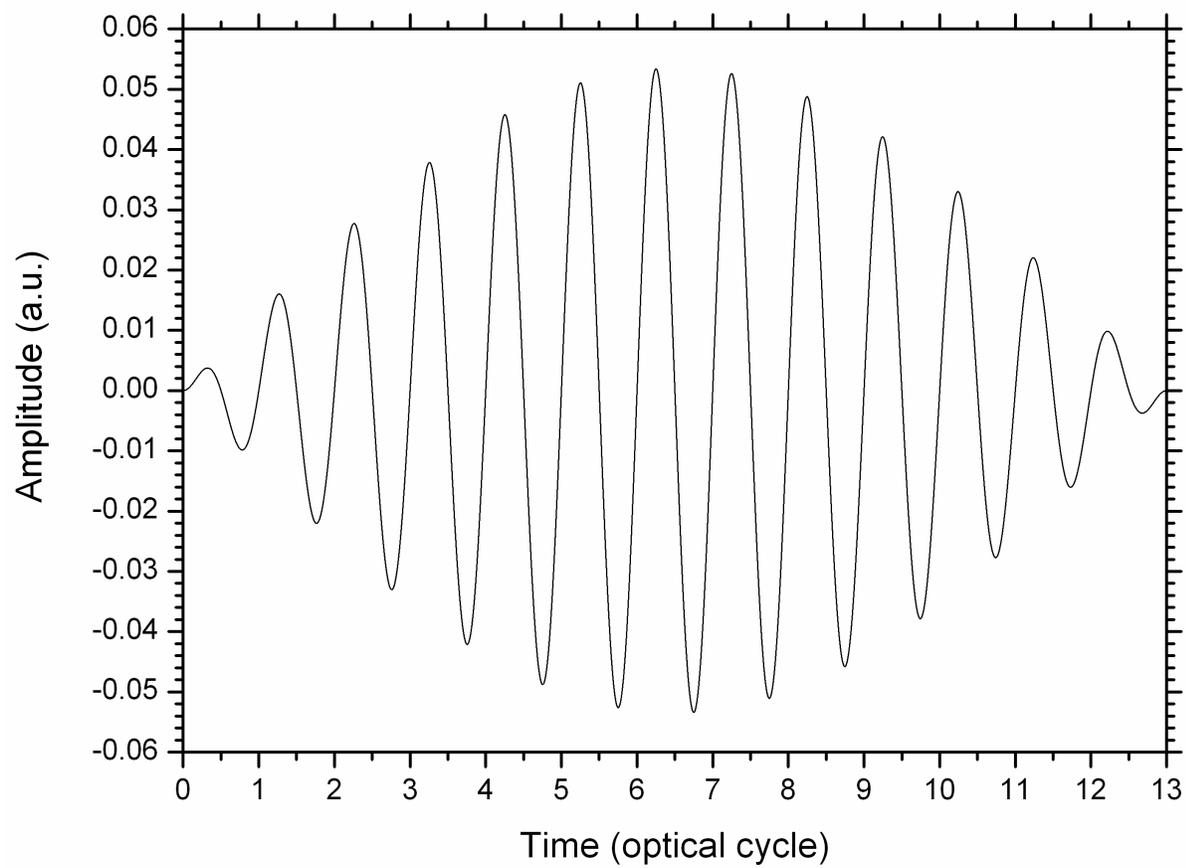} 
\caption{\label{fig:LASER} 
Electric field of the laser pulse adopted.} 
\end{figure}

The TDSE calculations and the RT-TDDFT calculations employing the adiabatic LDA \cite{LDAX,LDAC}, PBE \cite{PBE}, LB94 \cite{LB94}, and LFAs-PBE \cite{LFA} XC potentials, are performed 
with the program package \textsf{Octopus 4.0.1} \cite{Octo}. The system $\text{H}_{2}^{+}$ is described on a uniform real-space grid with a spacing of 0.19 bohr and a sphere of radius $r_m = 25$ 
bohr around the center of mass of the system. The nuclei are positioned parallel ($\theta = 0^{\circ}$) or perpendicular ($\theta = 90^{\circ}$) to the $x$-axis with a frozen equilibrium bond length of 
2.0 bohr \cite{Willock}. The electron-ion interaction is represented by norm-conserving Troullier-Martins pseudopotentials \cite{TM}. To obtain the HHG spectrum, $\text{H}_{2}^{+}$, which starts from 
the ground state, experiences an intense linearly polarized laser pulse at time $t=0$. The strong field interaction is generated by an oscillating electric field linearly polarized parallel to the $x$-axis: 
\begin{equation} 
v_{laser}({\bf{r}},t) = -x E_{0} \sin\left(\frac{\pi t}{T}\right) \sin\left(\omega_0 t\right). 
\label{eq:Vlx} 
\end{equation} 
Here the interaction with the electric field is treated in the dipole approximation and the length gauge \cite{Cormier}. The frequency of the laser pulse $\omega_0$ = 1.55 eV (i.e., the wavelength 
$\lambda = 800$ nm), the peak intensity $I_0 = 1 \times 10^{14}$ W/cm$^2$, and a pulse duration of 13 optical cycles (i.e., $T = 26 \pi / \omega_0 = 34.7$ fs) are adopted (see \Cref{fig:LASER}). 
To propagate the TDSE and TDKS equations, we adopt a time step of $\Delta t = 0.02$ a.u. (0.484 as) and run up to $t_f = 1500$ a.u. (36.3 fs), which corresponds to $7.5 \times 10^4$ time steps. 
The approximated enforced time-reversal symmetry (AETRS) algorithm is employed to numerically represent the time evolution operator \cite{aetrs}. 

In the HHG process, the electron released by tunnel ionization may travel far away from the simulation region ($r \leq r_m$). For computational efficiency, we separate the free-propagation space 
by employing the masking function \cite{Giovannini,Giovannini2}: 
\begin{equation} 
M(r) = \left\{\begin{array}{ll}& 1, \quad\ {r \leq r_m} \\ 
\\ 
& 1 - \sin^2(\frac{\pi}{2}\frac{r-r_m}{a}), \quad\ {r > r_m}. \end{array} \right. 
\label{eq:M} 
\end{equation} 
Here $a = 5$ bohr is the width of the masking function. At each time step, each Kohn-Sham (KS) orbital 
$\psi_j({\bf r},t)$ is multiplied by $M(r)$: 
\begin{equation} 
\psi_j({\bf r},t) \rightarrow M(\text{r}) \psi_j({\bf r},t). 
\label{eq:2} 
\end{equation} 
While the KS orbitals inside the simulation region are unaffected by the masking function, those at the boundary may gradually disappear and never return to the nuclei. Due to the masking function, 
the normalization of the KS orbitals may decrease in time. Therefore, the ionization probability of the KS orbital $\psi_j({\bf r},t)$ can be defined as 
\begin{equation} 
P_j(t) = 1 - \int |\psi_j({\bf r},t)|^2 d{\bf r}. 
\label{eq:Pj} 
\end{equation} 
After the electron density $\rho({\bf r},t) = \sum_{j=1}^{N} |\psi_j({\bf r},t)|^2$ is determined, the induced dipole moment is calculated by 
\begin{equation} 
d(t) = \int x \rho({\bf r},t) d{\bf r}, 
\label{eq:dipole} 
\end{equation} 
and the dipole acceleration is calculated using the Ehrenfest theorem \cite{Ehren}: 
\begin{equation} 
a(t) = \frac{d^{2}}{dt^{2}}d(t) = - \int \rho({\bf r},t) \frac{\partial v_{s}({\bf r},t)}{\partial x} d{\bf r}, 
\label{eq:Ehren} 
\end{equation} 
where $v_{s}({\bf r},t)$ is the time-dependent KS potential. The HHG spectrum is calculated by taking the Fourier transform of the dipole acceleration \cite{HHG_eval}: 
\begin{equation} 
H(\omega) = {\left | \frac{1}{t_f} \int_{0}^{t_f} a(t) e^{-i\omega t} dt \right |}^{2}. 
\label{eq:Hw} 
\end{equation} 

As the asymptotic behavior of the XC potential can be important for a wide variety of time-dependent properties, in this work, we examine the ionization probability of HOMO ($1{\sigma_g}$) 
[Eq.\ (\ref{eq:Pj})], the induced dipole moment [Eq.\ (\ref{eq:dipole})], and the HHG spectrum [Eq.\ (\ref{eq:Hw})] for $\text{H}_{2}^{+}$ aligned parallel ($\theta = 0^{\circ}$) or perpendicular 
($\theta = 90^{\circ}$) to the laser polarization, using the exact TDSE and RT-TDDFT with the adiabatic LDA, PBE, LB94, and LFAs-PBE XC potentials.

\section{Results and Discussion} 

The vertical ionization potential of a neutral molecule, defined as the energy difference between the cationic and neutral charge states, is identical to the minus HOMO energy of the neutral 
molecule obtained from the exact KS potential \cite{Janak,Fractional,HOMO2,Levy84,1overR,HOMO}. Accordingly, the minus HOMO energy calculated using approximate KS-DFT has been one 
of the most common measures of the quality of the underlying XC potential \cite{LCAC,EB,LFA,DD}. As shown in \Cref{table:HOMO}, the minus HOMO ($1{\sigma_g}$) energy of the one-electron 
$\text{H}_{2}^{+}$ system, obtained with the exact time-independent Schr\"odinger equation (TISE) is 29.96 eV. However, the minus HOMO energies of $\text{H}_{2}^{+}$, calculated using the 
LDA and PBE XC potentials are about 6 eV smaller than the exact value, as the LDA or PBE XC potential exhibits an exponential decay (not the correct $(-1/r)$ decay) in the asymptotic region 
($r \rightarrow \infty$) of a molecule. By contrast, the minus HOMO energies of $\text{H}_{2}^{+}$, calculated using the LB94 and LFAs-PBE XC potentials are in good agreement with the exact 
value (within an error of 1.8 eV), due to the correct asymptotic behavior of the LB94 and LFAs-PBE XC potentials. 

\begin{figure} 
\includegraphics[width=1.0\textwidth]{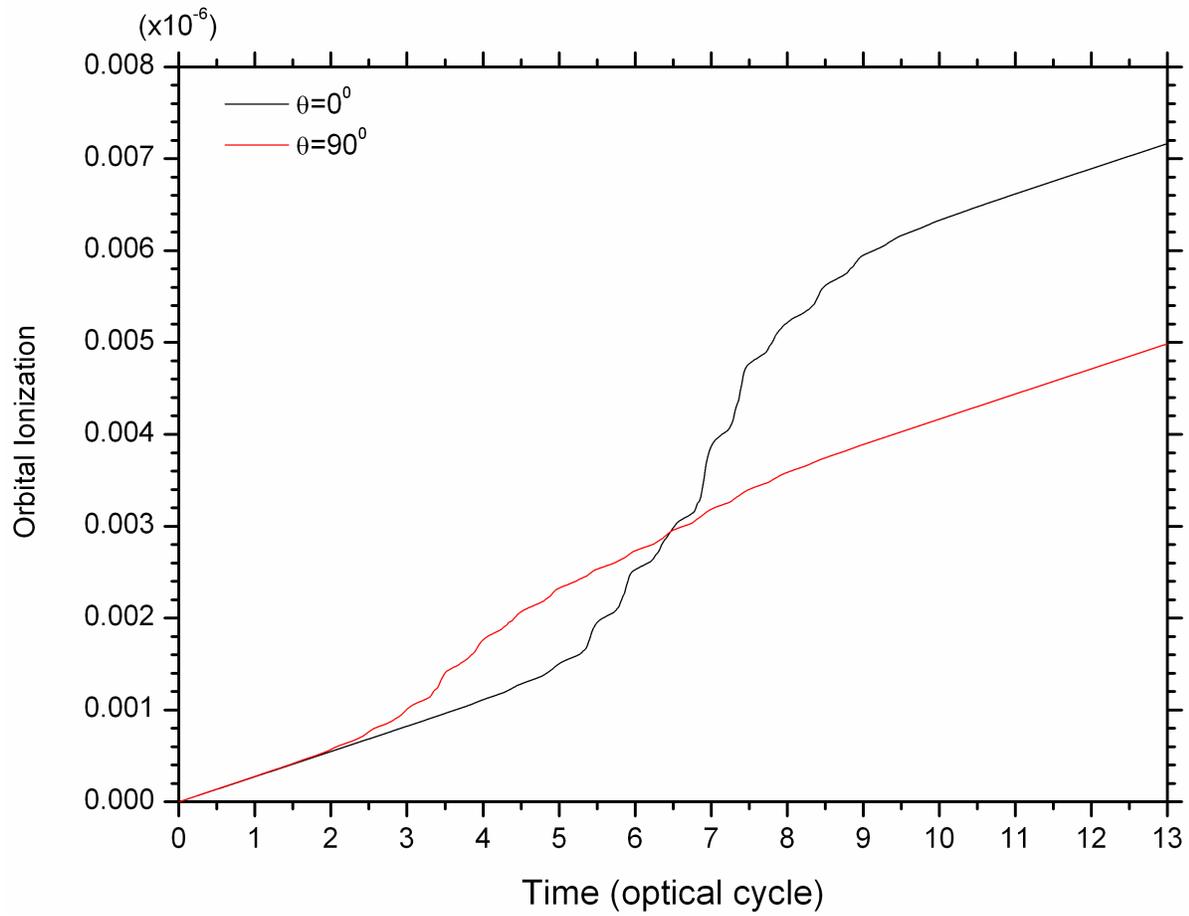} 
\caption{\label{fig:H2+-orbital-occu-XY} 
Ionization probability of HOMO ($1{\sigma_g}$) for $\text{H}_{2}^{+}$ aligned parallel ($\theta = 0^{\circ}$) or perpendicular ($\theta = 90^{\circ}$) to the laser polarization, 
calculated using the exact TDSE.} 
\end{figure}

\begin{figure} 
\includegraphics[width=1.0\textwidth]{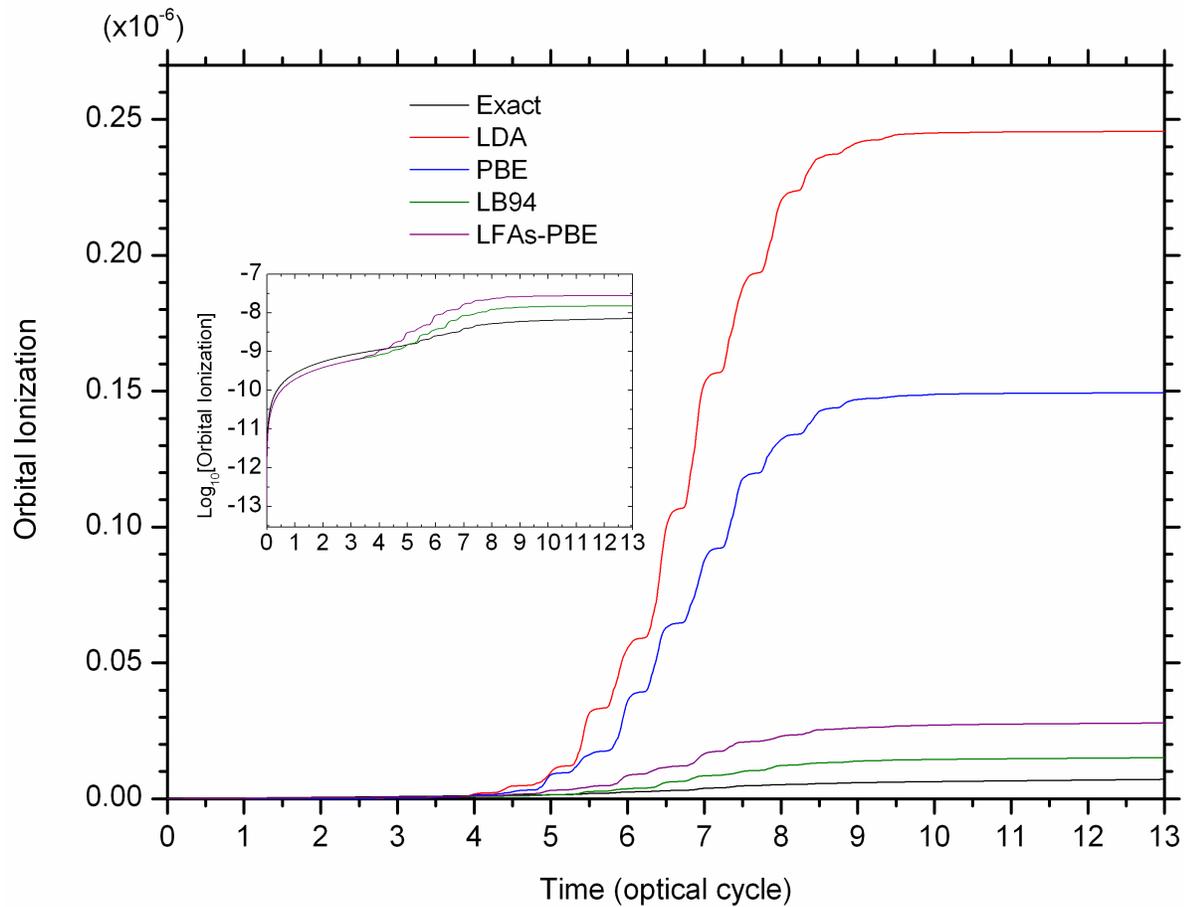} 
\caption{\label{fig:H2+-orbital-occu-x-1} 
Ionization probability of HOMO ($1{\sigma_g}$) for $\text{H}_{2}^{+}$ aligned parallel ($\theta = 0^{\circ}$) to the laser polarization, calculated using the exact TDSE 
and RT-TDDFT with various adiabatic XC potentials. Inset shows the TDSE, LB94, and LFAs-PBE results on a logarithmic scale.} 
\end{figure}

\begin{figure} 
\includegraphics[width=1.0\textwidth]{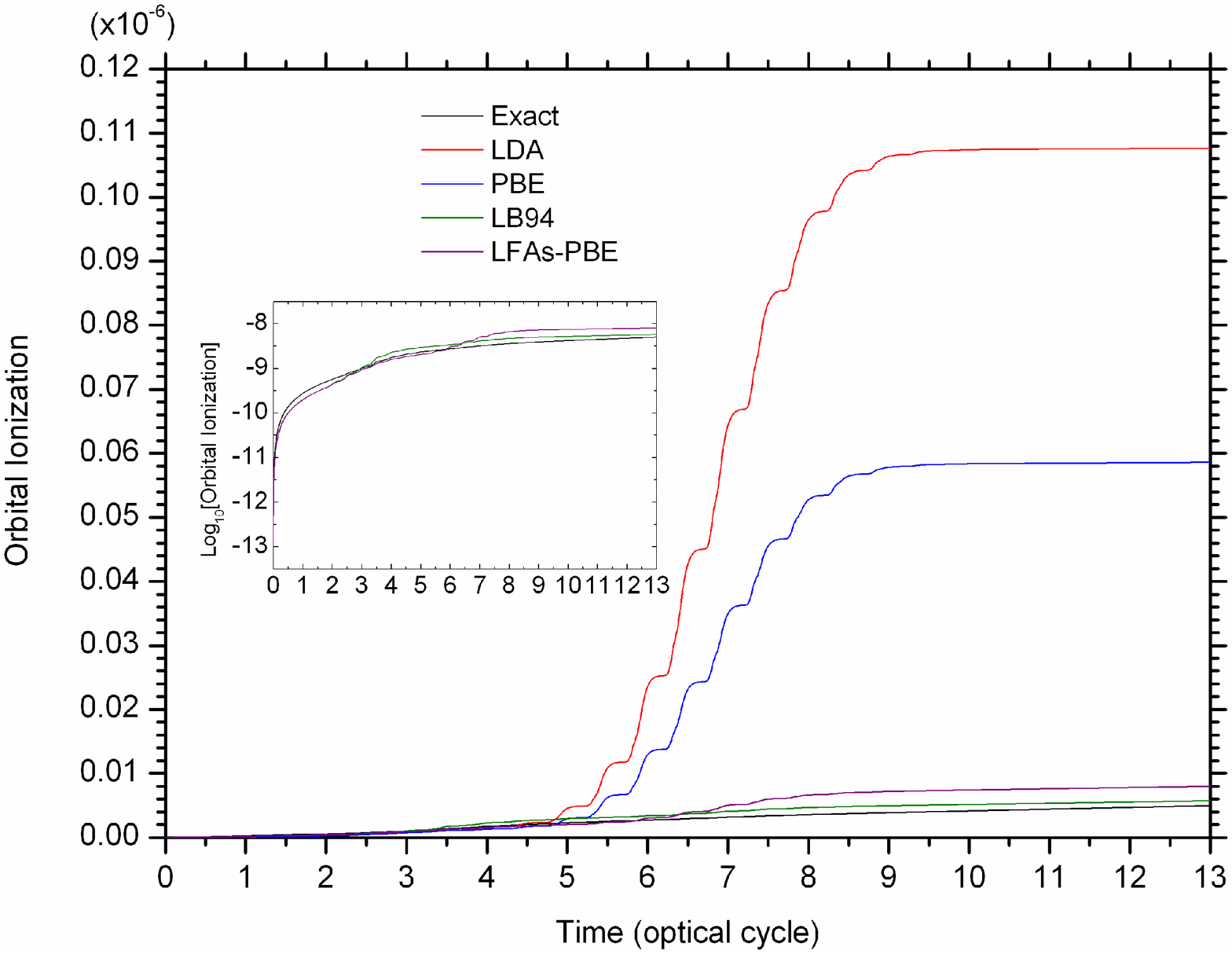} 
\caption{\label{fig:H2+-orbital-occu-Y-1} 
Same as Fig.\ \ref{fig:H2+-orbital-occu-x-1}, but for $\text{H}_{2}^{+}$ aligned perpendicular ($\theta = 90^{\circ}$) to the laser polarization.} 
\end{figure}

\Cref{fig:H2+-orbital-occu-XY} plots the ionization probability of HOMO ($1{\sigma_g}$) for $\text{H}_{2}^{+}$ aligned parallel ($\theta = 0^{\circ}$) or perpendicular ($\theta = 90^{\circ}$) to the 
laser polarization, calculated using the exact TDSE. After 5 optical cycles, the ionization probabilities of HOMO calculated using the LB94 and LFAs-PBE XC potentials become much more accurate 
than those calculated using the LDA and PBE XC potentials for both the parallel (\Cref{fig:H2+-orbital-occu-x-1}) and perpendicular (\Cref{fig:H2+-orbital-occu-Y-1}) alignments. Owing to the correct 
$(-1/r)$ asymptote, the LB94 and LFAs-PBE XC potentials are more difficult to ionize as compared with the LDA and PBE XC potentials (which decay much faster in the asymptotic region). 

\begin{figure}
\includegraphics[width=1.0\textwidth]{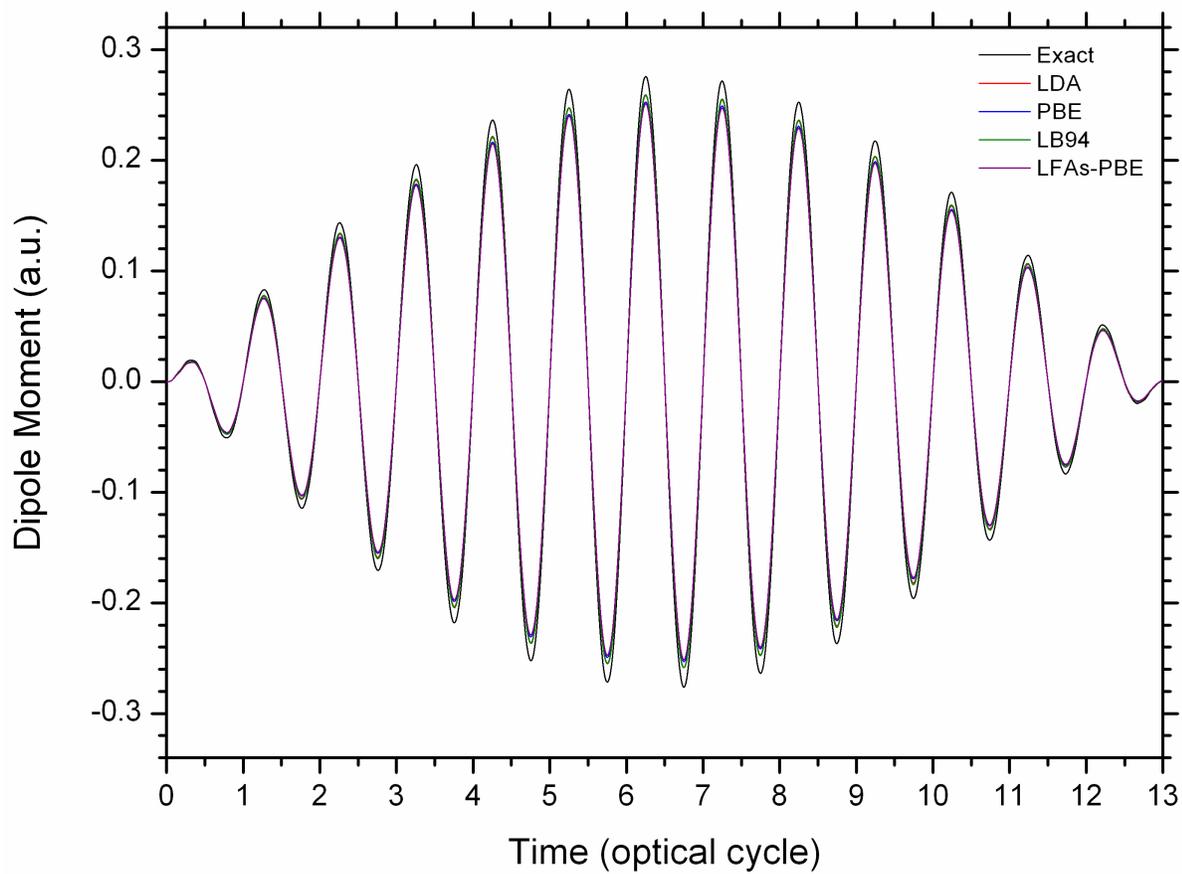}
\caption{\label{fig:H2+-Dipole-x} 
Induced dipole moment for $\text{H}_{2}^{+}$ aligned parallel ($\theta = 0^{\circ}$) to the laser polarization, calculated using the exact TDSE 
and RT-TDDFT with various adiabatic XC potentials.} 
\end{figure}

\begin{figure}
\includegraphics[width=1.0\textwidth]{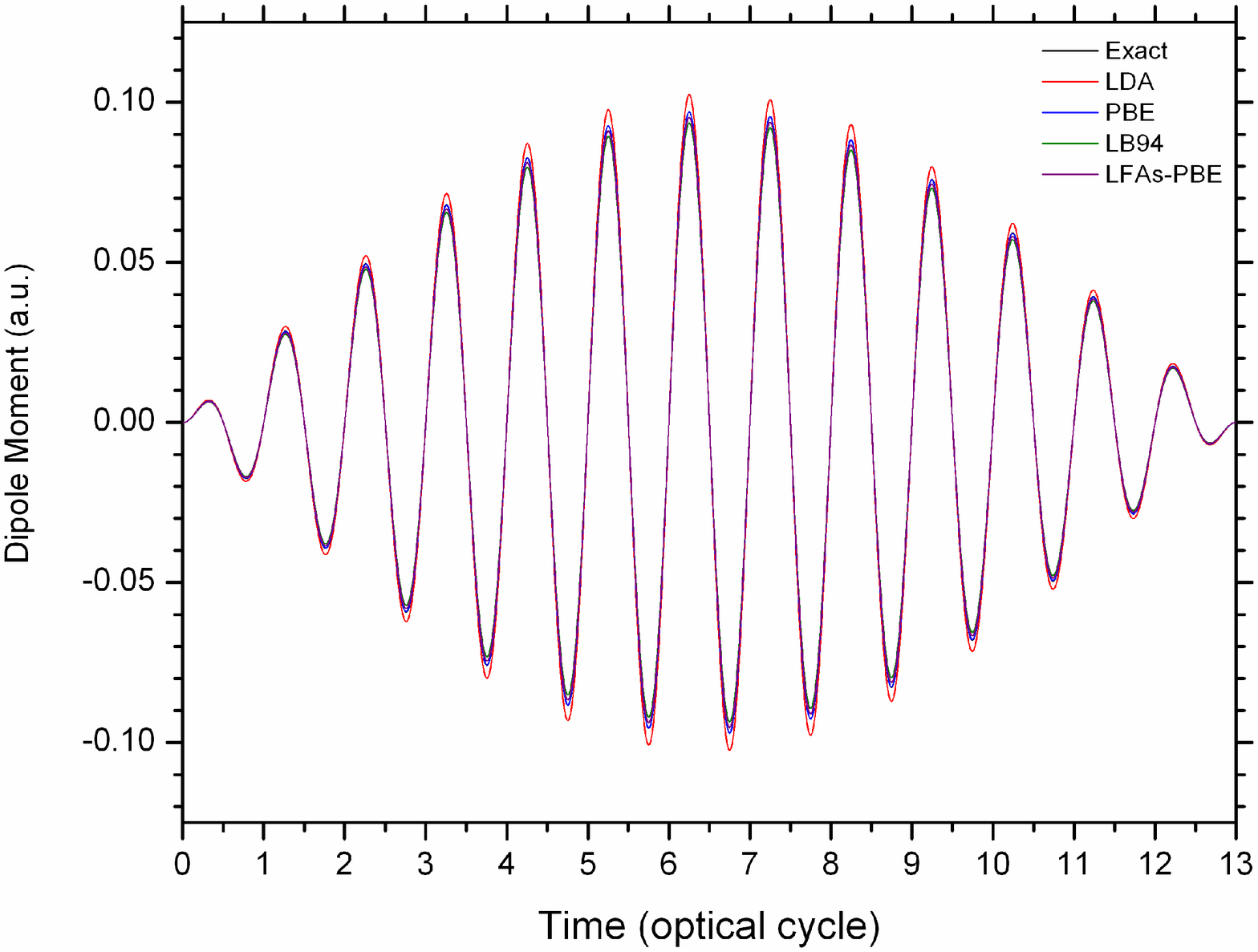} 
\caption{\label{fig:H2+-Dipole-y} 
Same as Fig.\ \ref{fig:H2+-Dipole-x}, but for $\text{H}_{2}^{+}$ aligned perpendicular ($\theta = 90^{\circ}$) to the laser polarization.} 
\end{figure}

In addition, we examine the induced dipole moment for $\text{H}_{2}^{+}$ aligned parallel (\Cref{fig:H2+-Dipole-x}) or perpendicular (\Cref{fig:H2+-Dipole-y}) to the laser polarization, calculated 
using the exact TDSE and RT-TDDFT with various adiabatic XC potentials. The induced dipole moments obtained with the adiabatic XC potentials in RT-TDDFT are very similar, and are all close to 
those obtained with the exact TDSE, revealing that the choice of the XC potential has only a minor effect on the induced dipole moments. 

\begin{figure}
\includegraphics[width=1.0\textwidth]{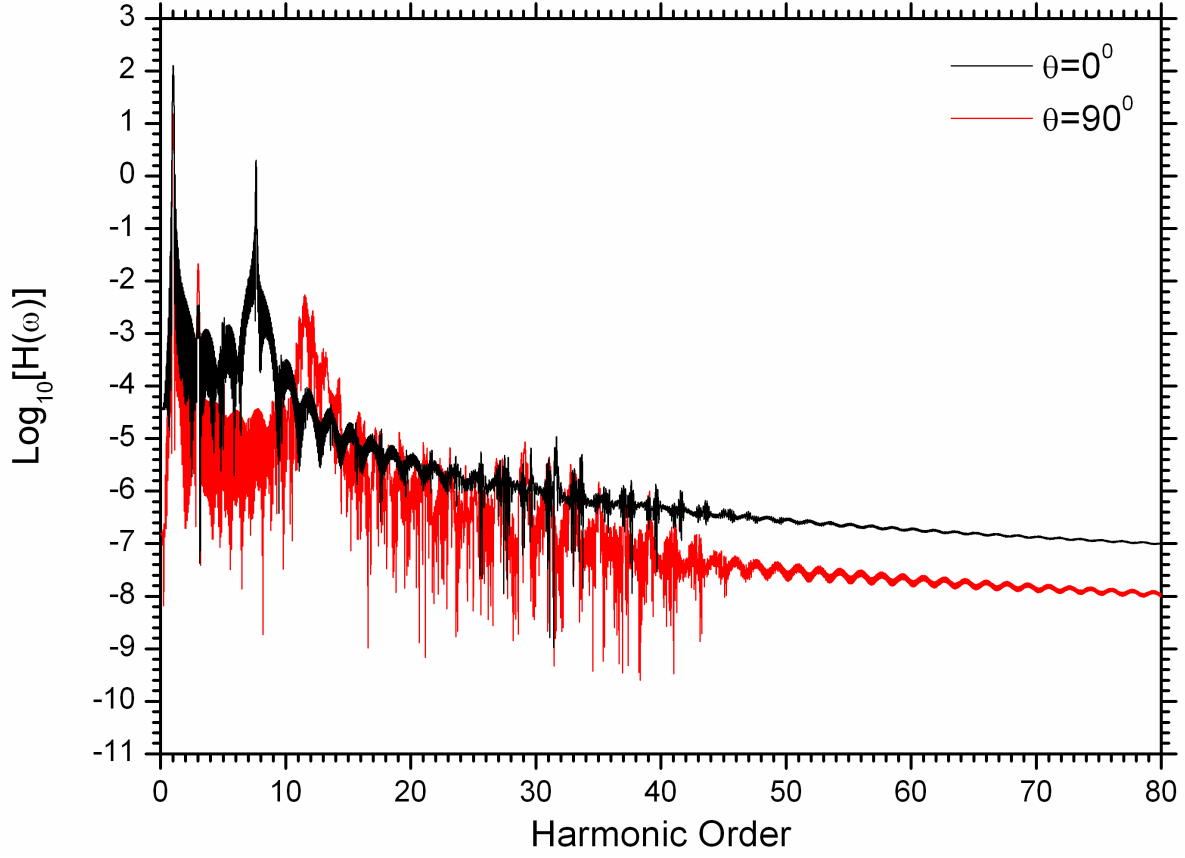} 
\caption{\label{fig:H2+-HHG-XY} 
HHG spectrum for $\text{H}_{2}^{+}$ aligned parallel ($\theta = 0^{\circ}$) or perpendicular ($\theta = 90^{\circ}$) to the laser polarization, calculated using the exact TDSE. 
Here the unit of H($\omega$) is \AA/($\hbar{\text{eV}}^{2}$).} 
\end{figure}

\begin{figure}
\includegraphics[width=1.0\textwidth]{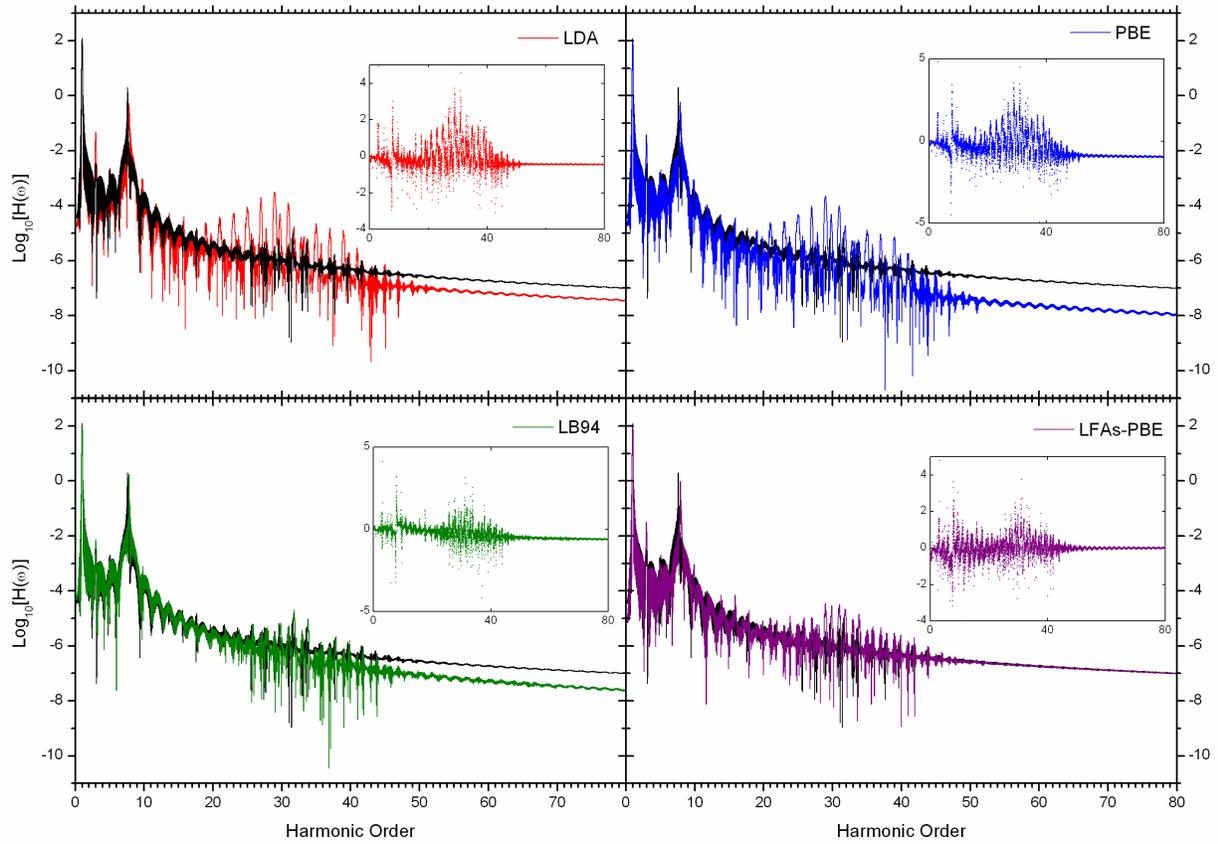}
\caption{\label{fig:H2+-HHG-1} 
HHG spectrum for $\text{H}_{2}^{+}$ aligned parallel ($\theta = 0^{\circ}$) to the laser polarization, calculated using the exact TDSE (black curve) and RT-TDDFT with 
various adiabatic XC potentials. Insets show the differences between the RT-TDDFT and TDSE results. Here the unit of H($\omega$) is \AA/($\hbar{\text{eV}}^{2}$).} 
\end{figure}

\begin{figure} 
\includegraphics[width=1.0\textwidth]{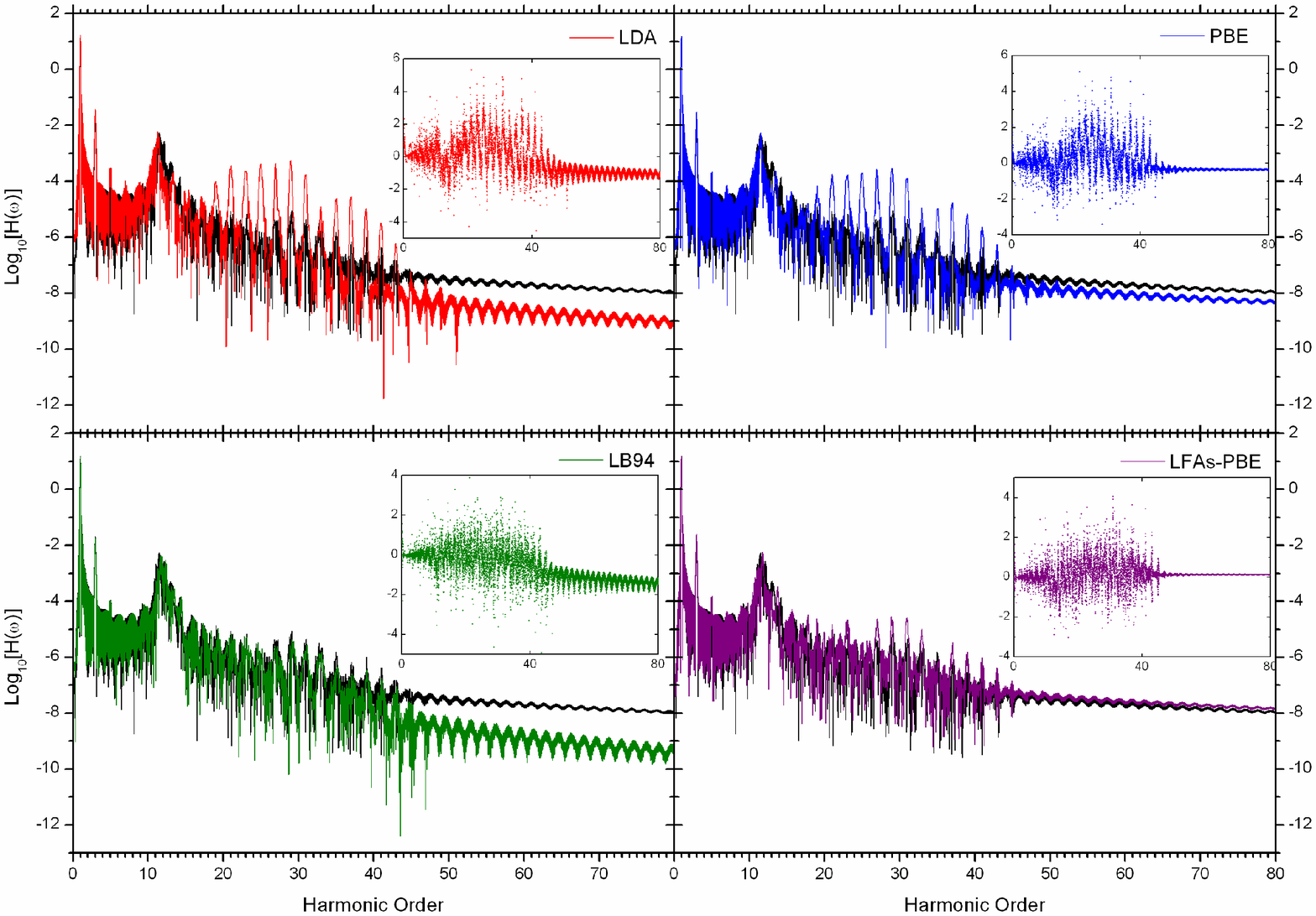} 
\caption{\label{fig:H2+-HHG-Y-1} 
Same as Fig.\ \ref{fig:H2+-HHG-1}, but for $\text{H}_{2}^{+}$ aligned perpendicular ($\theta = 90^{\circ}$) to the laser polarization.} 
\end{figure}

In comparison to the HHG spectra obtained with the exact TDSE (\Cref{fig:H2+-HHG-XY}), we plot the HHG spectrum for $\text{H}_{2}^{+}$ aligned parallel (\Cref{fig:H2+-HHG-1}) or perpendicular 
(\Cref{fig:H2+-HHG-Y-1}) to the laser polarization, calculated using various adiabatic XC potentials in RT-TDDFT. In contrast to the work of Wasserman and co-workers \cite{Wasserman}, the 
LFAs-PBE potential, which is an AC model potential, performs significantly better than the LDA, PBE, and LB94 XC potentials, especially for the higher harmonics (from the 50$^{th}$ to 80$^{th}$ 
harmonics). As a result, the fine details of the XC potential (not just the XC potential asymptote) are expected to be important for the calculated HHG spectra.

\section{Conclusions} 

In summary, we have assessed the performance of our recently proposed LFAs-PBE XC potential for the HHG spectra and related properties of $\text{H}_{2}^{+}$ aligned parallel and 
perpendicular to the polarization of an intense linearly polarized laser pulse, using adiabatic RT-TDDFT. The results have been compared with those obtained with the exact TDSE as well as 
those calculated using adiabatic RT-TDDFT with the LDA, PBE, and LB94 XC potentials. Due to the correct $(-1/r)$ asymptote, both the LB94 and LFAs-PBE XC potentials significantly 
outperform the conventional LDA and PBE XC potentials for the properties sensitive to the XC potential asymptote. For the HHG spectra, the LFAs-PBE potential performs significantly better 
than the LDA, PBE, and LB94 potentials, implying that the fine details of the XC potential (not just the XC potential asymptote) could also be important for HHG spectra. Relative to the LB94 
potential and other AC model potentials that are not functional derivatives, the LFAs-PBE potential has been shown to be significantly superior in performance for ground-state energies and 
related properties \cite{LFA}. Therefore, the LFAs-PBE potential, which has a computational cost similar to that of the popular PBE potential, can be very promising for studying the ground-state, 
excited-state, and time-dependent properties of large electronic systems, extending the applicability of density functional methods for a diverse range of applications.

\begin{acknowledgments} 

This work was supported by the Ministry of Science and Technology of Taiwan (Grant No.\ MOST104-2628-M-002-011-MY3), National Taiwan University (Grant No.\ NTU-CDP-105R7818), 
the Center for Quantum Science and Engineering at NTU (Subproject Nos.:\ NTU-ERP-105R891401 and NTU-ERP-105R891403), and the National Center for Theoretical Sciences of Taiwan. 

\end{acknowledgments}

\newpage 
\begin{table} 
\begin{ruledtabular} 
\caption{\label{table:HOMO} 
Minus HOMO ($1{\sigma_g}$) energy (in eV) of $\text{H}_{2}^{+}$, calculated using the exact TISE and KS-DFT with various XC potentials.} 
\begin{tabular}{cccccc} 
Orbital & Exact & LDA & PBE & LB94 & LFAs-PBE \\ 
\hline 
$1\sigma_g$ & 29.96 & 23.24 & 23.72 & 28.24 & 28.25 \\ 
\end{tabular} 
\end{ruledtabular} 
\end{table} 

\newpage 

\newpage 

\newpage 

\newpage 

\newpage

\newpage

\newpage

\newpage

\newpage 

\end{document}